\documentclass[lettersize,journal]{IEEEtran}
\usepackage{amsmath,amsfonts}
\usepackage{algorithmic}
\usepackage{algorithm}
\usepackage{array}
\usepackage[caption=false,font=normalsize,labelfont=sf,textfont=sf]{subfig}
\usepackage{textcomp}
\usepackage{stfloats}
\usepackage{url}
\usepackage{verbatim}
\usepackage{graphicx}
\usepackage{cite}
\usepackage{bm}
\usepackage{xcolor}
\usepackage{multirow}
\begin{document}

\newcommand{\mi}[1]{\textcolor{blue}{#1}}  
\newcommand{\am}[1]{\textcolor{red}{[AM: #1]}} 
\newcommand{\aj}[1]{\textcolor{magenta}{#1}} 
\newcommand{\ra}[1]{\textcolor{green}{#1}} 
\newcommand{\scheme}{\textsc{Pai}} 
\newcommand{\rev}[1]{\textcolor{black}{#1}} 

\title{\textsc{Pai}: Fast, Accurate, and Full Benchmark Performance Projection with AI}
\vspace{-0.4in}

\author{Avery Johnson, Mohammad Majharul Islam, Riad Akram, and Abdullah Muzahid

\thanks{Manuscript received March 19, 2025; revised June 11, 2025.}
\thanks{Avery Johnson and Abdullah Muzahid are with Texas A\&M University, College Station, TX 77840 USA. (e-mail: averyjohnson@tamu.edu, abdullah.muzahid@tamu.edu). }
\thanks{Mohammad Majharul Islam and Riad Akram are with the Intel Corporation, Hillsboro, OR 97124 USA. (e-mail: mohammad.majharul.islam@intel.com, riad.akram@intel.com).}
}



\maketitle
\begin{abstract}
The exponential increase in complex IPs within modern SoCs, driven by Moore’s Law, has created a pressing need for fast and accurate hardware-software power-performance analysis. 
Traditional performance simulators (such as cycle accurate simulators) are often too slow to simulate full benchmarks within a reasonable timeframe; require considerable effort for development, maintenance, and extensions; 
and are prone to errors, 
making pre-silicon performance projections and competitive analysis increasingly challenging. Prior attempts in addressing this challenge using machine learning fall short as they are either slow, inaccurate or unable to predict the performance of full benchmarks. To address these limitations, we present \scheme, the {\em first} technique to accurately predict full benchmark performance without relying on detailed simulation or instruction-wise encoding. At the heart of \scheme\ is a hierarchical Long Short Term Memory (LSTM)-based model that takes a trace of microarchitecture independent features from a program execution and predicts performance metrics. We present the detailed design, implementation and evaluation of \scheme. Our initial experiments showed that \scheme\ can achieve an average IPC prediction error of 
\rev{9.35\%} for SPEC CPU 2017 benchmark suite while taking only 2 min 57 sec for the entire suite. This prediction error is comparable to prior state-of-the-art techniques while requiring 3 orders of magnitude less time.

\end{abstract}


\vspace{-0.2in}
\section{Introduction}
\label{sec-intro}




\IEEEPARstart{E}{mergence} of Internet-of-Things and smart devices are making System-on-Chips (SoCs) a popular design option for their size, cost, and performance efficiency. Modern SoCs are sophisticated, incorporating a wide range of Intellectual Property (IP) blocks such as CPUs, GPUs, memory subsystems, various hardware accelerators, communication interfaces, sensors, etc. Furthermore, the IPs themselves are becoming more and more complex. This necessitates any tool that can make designing SoCs faster and easier to do. 


Performance modeling is an integral part for designing the next generations SoCs. It is a way to determine what design choices are better and therefore, should be considered at different stages of the design cycle. However, the complexity of modern SoCs make it a challenging task to model or predict performance. Traditionally, performance modeling is done through an architectural performance simulator such as Z-sim~\cite{zsim}, Gem5~\cite{gem5}, ChampSim~\cite{champsim}, etc. These simulators simulate varying degrees of architectural events in a cycle-by-cycle fashion, eventually leading to a set of desired metrics such as Instruction Per Cycle (IPC), cache misses, branch mispredictions, power consumption, etc. The simulators, although quite accurate, are extremely slow.
They operate at Kilo Instructions per Second (KIPS) speed and can take from few hours to days or weeks to simulate a full benchmark. As an example, it takes 8 hours to simulate
a single core for one second of execution in Zsim~\cite{zsim}. This exorbitant runtime puts a serious hurdle to the performance estimation step 
of the SoC design process. As every iteration of the design cycle needs to simulate many benchmarks to evaluate and eventually, improve the design,
these simulators impose a significant development cost and delay. 
Furthermore, even state-of-the-art simulators aren't perfect; with an average error of 9.7\% and 13\% respectively for Z-sim and Gem5 \cite{simnet}.
What is needed is a {\em performance modeling/prediction technique that is both fast and accurate}. So, we ask ourselves the following question: Can this problem be solved with emergent machine learning models?


Given that machine learning models are fast at inference (compared to the time for cycle-accurate simulation), it's no surprise that it has already been applied 
in earlier research~\cite{li24_sc, tao, munigala, barboza}. Most recently, Li et al.~\cite{li24_sc} propose to combine two deep learning models - one to capture program representations by encoding individual instructions and one to capture microarachitectural specifications. The third model combines the output of those two models and predicts performance of a program. As the model needs to encode individual instructions, it's practicality is limited by how many instructions can be processed before the model starts to exhibit poor prediction accuracy. TAO~\cite{tao} uses a dataset of microarchitecture independent features 
from a detailed simulator. This dataset is used to train a multi-head attention network in order to predict IPC, branch mispredictions, etc.
TAO requires a detailed simulator for the the dataset to begin with. Therefore, it is inherently limited by the speed and accuracy of detailed simulator.
Munigala et al.~\cite{munigala} use a multilayer perceptron 
to predict the bandwidth and latency of a benchmark based on seven architectural parameters (core count, clock frequency, Last Level Cache (LLC) size, etc.). 
However, like TAO, the performance metrics are obtained from a detailed simulator and hence, it faces similar issues.
Barboza et al.~\cite{barboza} use many performance counters across multiple benchmarks to train a classifier to predict performance bottlenecks. The goal is to find performance degrading components of a microarchitecture. Like earlier works, it also relies on detailed simulation and considers only a single architecture. Besides these techniques, there are earlier attempts to use machine learning for performance prediction~\cite{ipek06,li16, xapp}. However, they are program and/or architecture specific, and therefore, lacking generality. 

To overcome the limitations of prior work and propose a general approach capable of full benchmark performance prediction, we start by setting three design goals - {\em (i) detailed simulation independence, (ii) microarchitecture independence, and (iii) full benchmark prediction capability}. With these design goals in mind, we propose \scheme. \scheme\ is the {\em first} technique to accurately predict full benchmark performance for seen and unseen programs and architectures without relying on detailed simulation or instruction-wise encoding. At the heart of \scheme\ is a deep learning model based on hierarchical LSTM~\cite{lstm}. The model takes a trace of microarchitecture independent features from a program execution. The features are collected either from real machines or instruction set emulators (such as Simics~\cite{simics} that is orders of magnitude faster than detailed simulation). \scheme\ augments those features with high level microarchitectural specifications and provides performance prediction of the full benchmark both for seen and unseen programs and architectures. We provide a detailed design of \scheme's deep learning model as well as the approach to generate necessary dataset for training the model.
Our initial experiments showed that \scheme\ can achieve an average IPC prediction error of 9.35\% for SPEC CPU 2017 benchmark suite while taking only 2 min 57 sec for the entire suite. This prediction error is comparable to prior state-of-the-art techniques~\cite{tao,li24_sc} while requiring 3 orders of magnitude less time.



\vspace{-0.2in}
\section{Main Idea: PAI}
\label{sec-pai}
\subsection{Overview of \scheme}
\label{sec-overview}
\begin{figure}[htpb]
\centering
\includegraphics[width=0.55\columnwidth,]{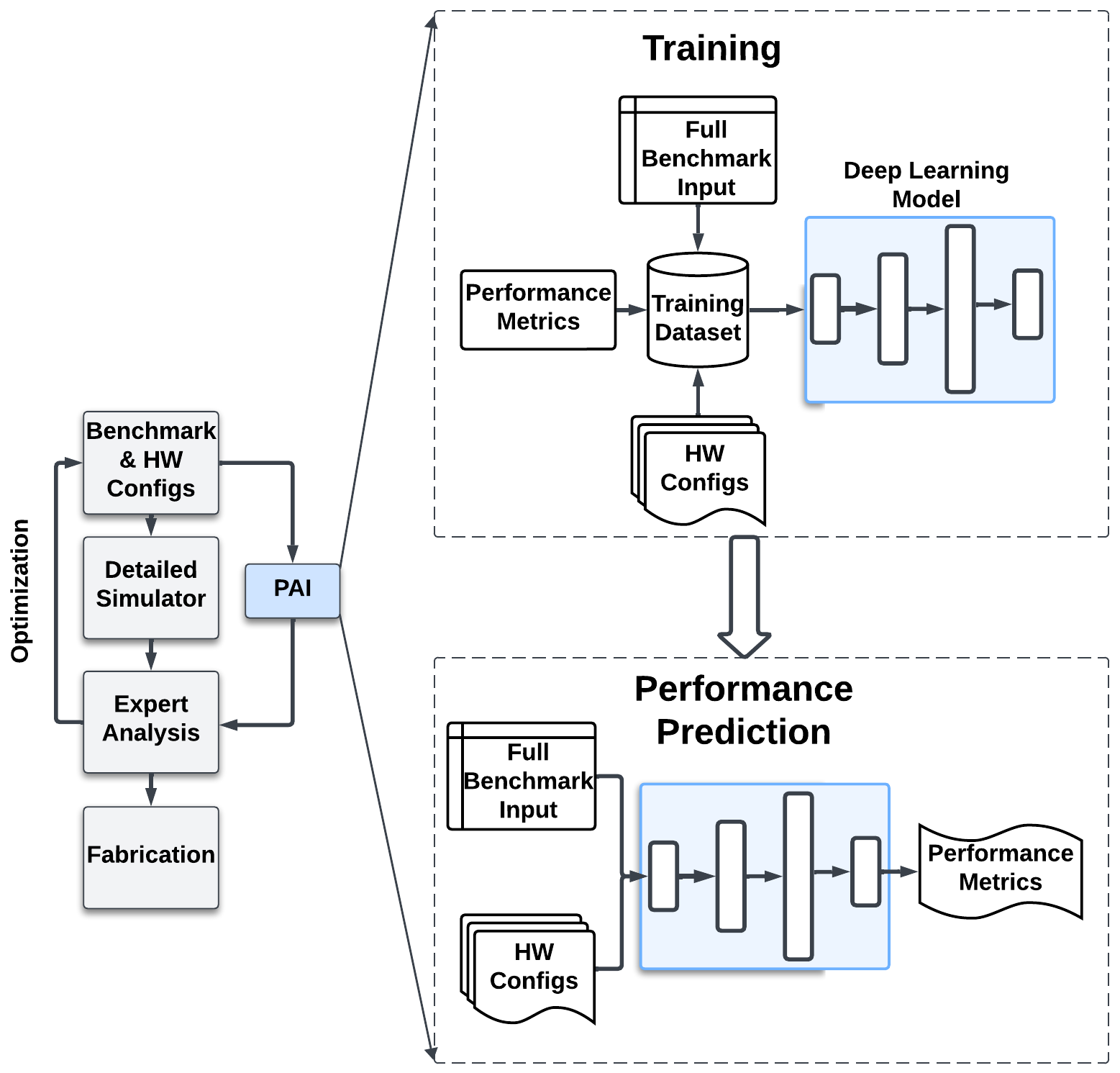}%
\caption{Overview of how \scheme\ works by augmenting the design cycle.}
\label{fig_overview}
\end{figure}

At a high level, \scheme\ works by augmenting the traditional design cycle with an additional deep learning-based predictor (as shown in Figure~\ref{fig_overview}) that provides a first-order approximation of performance metrics. The intuition is that by approximating the performance potential of different design choices for seen and unseen programs, \scheme\ enables hardware designers to disregard many of those choices and focus on only a few {\em good} ones. To achieve this goal, \scheme\ works in two phases. In the training phase, a dataset comprising of hardware configurations, execution traces of microarchitecture independent features, and performance metrics is collected for entire benchmarks using native machines or ISA emulators. Since no detailed simulators are used for this phase, \scheme\ can rely on a dataset large enough to train any deep learning model. In the prediction phase, the trained model is provided with execution traces and hardware configurations as inputs. The \scheme\ model iterates over a trace of inputs and predicts performance metrics which are then summed up to provide full benchmark performance. By abstracting the hardware configurations and application behavior in a trace, \scheme\ can predict performance for any hardware configuration and program - seen or unseen.

\begin{figure}[htpb]
\centering
\includegraphics[width=0.65\columnwidth]{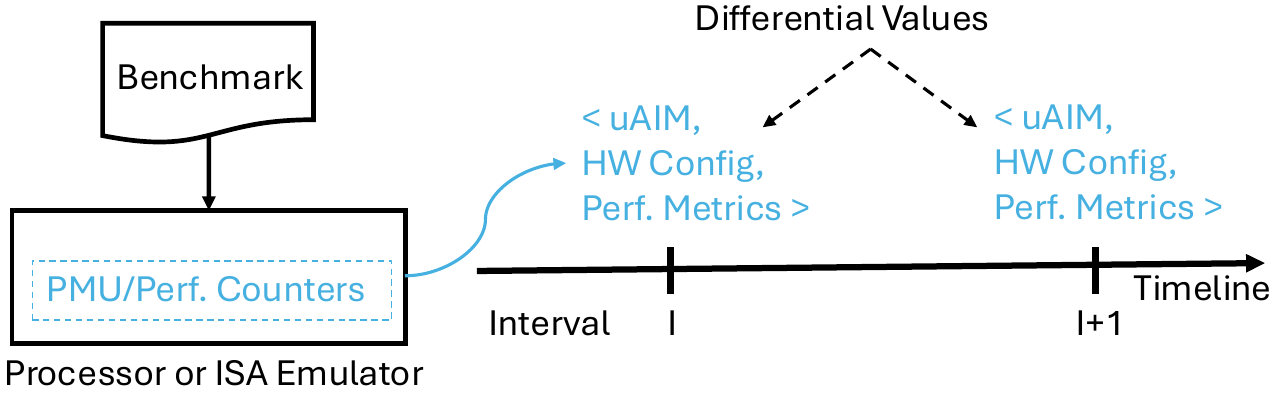}%
\caption{Overview of how \scheme\ collects data for training.}
\label{fig_data_collection}
\vspace{-0.4cm}
\end{figure}

\subsection{Dataset Construction}
\label{sec-training-data}
For our training dataset, we need to capture the benchmark behavior, hardware configurations, and performance metrics. To capture the benchmark behavior, we use normalized microarchitecture independent metrics (uAIMs). Those can be captured from various hardware performance counters from native machines or ISA emulators such as Simics~\cite{simics}. Intel provides a Performance Monitoring Unit (PMU) to capture various core and uncore metrics~\cite{intel_pmu}. Similarly, Simics provides a diverse range of uAIMs to capture the current state of program execution. Examples of uAIMs are number of taken/not taken branches, page faults, memory reads/writes, cache misses at different levels, number of different instructions, etc. In order to capture long running benchmarks or entire benchmarks, a snapshot of uAIMs is recorded at a certain interval (e.g., every 10M instructions). Each such snapshot consists of differential values between the current and previous counters (as shown in Figure~\ref{fig_data_collection}). Each snapshot is associated with hardware configurations and performance metrics (during the period from the last snapshot to the current one). Examples of hardware configurations are core count, thread count, size of different levels of cache, clock frequency, instruction issue width, ROB size, etc. Examples of performance metrics are Instruction Per Cycle (IPC), number of clock cycles, etc.


\begin{figure}[htpb]
\centering
\begin{minipage}{0.45\columnwidth}
\centering
\includegraphics[width=0.7\columnwidth]{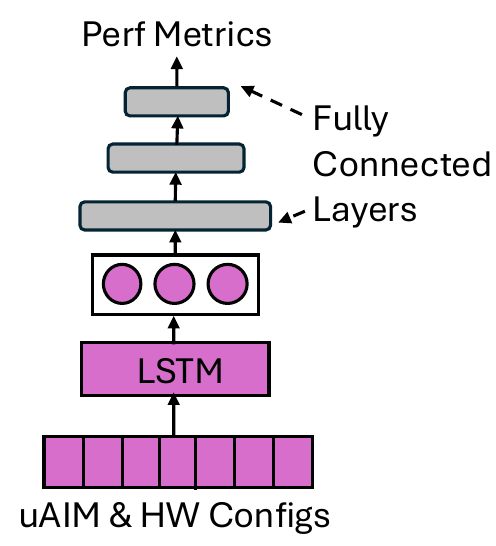}
 \caption{Simple LSTM model.}
\label{simple-model}
\end{minipage}
\hfill
\begin{minipage}{0.48\columnwidth}
\centering
\includegraphics[width=0.9\columnwidth]{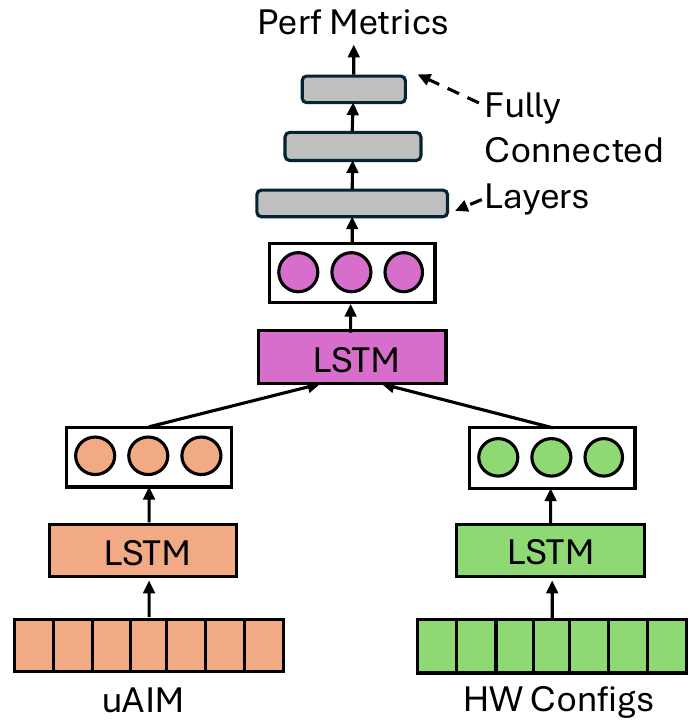}
\caption{Hierarchical LSTM model.}
\label{hier-model}
\end{minipage}
 \vspace{-0.2cm}
\label{fig-lstm}
\end{figure}

\vspace{-0.2in}
\subsection{Model Development} \label{development}
\label{sec-model}
Predicting performance metrics such as IPC can be considered as a regression problem, so we started with a linear regression model and a simple multilayer perceptron model. The linear regression model is too simplistic to capture the non-linear relation between the input and the output. Similarly, a multilayer perceptron is not sophisticated enough to achieve a reasonably good accuracy for a dataset that has non-linear and sequential patterns (Section~\ref{sec-model-analysis}). So, we turned our focus to LSTM that is designed for sequence learning~\cite{lstm}.


We experimented with two LSTM-based designs - Simple and Hierarchical. Figure~\ref{simple-model} shows the simple model. It combines the uAIMs and hardware configurations to an LSTM layer which produces some latent representation of the inputs. The latent vector is passed through one or more fully connected layers to produce the final IPC prediction. Figure~\ref{hier-model} shows the hierarchical LSTM model. One LSTM layer takes on uAIMs whereas another LSTM layer takes on hardware configurations. The resultant latent vectors are processed through a second level LSTM layer to produce the final latent vector. This latent vector is then passed through one or more fully connected layers (just like the simple design) to produce the final output.

\section{Evaluation}
\label{sec-eval}

\subsection{Experimental Setup}
\label{sec-setup}
To evaluate \scheme\ we used the SPEC CPU2017~\cite{speccpu17} benchmark suite due to its representativeness. We developed \scheme's training and testing framework using PyTorch and evaluated its performance on an Intel® Xeon® Silver 6336Y 24-core CPU. 


\subsection{Dataset Preparation}
\label{sec-dataset}
To prepare the training and testing datasets, we used Simics to collect 128 uAIMs every 10 million instructions during program execution. 
For each interval, we retrieved IPCs from 15 different configurations (SKUs) of Intel® Xeon® 6 processors, capturing performance variations due to different clock speeds, core counts, cache sizes, etc. Each datapoint in the dataset consists of a snapshot of these uAIMs (differential values) and the corresponding IPCs. Overall, the generated dataset consists of 1.3 million datapoints. 
A breakdown of the different uAIM features is shown in Table~\ref{features_breakdown}.

\begin{table}[ht]
\caption{Breakdown of uAIM features\label{tab:table1}}
\centering
\scalebox{0.7}{
\begin{tabular}{||p{2cm}||p{4.5cm}||p{0.75cm}||}
\hline\hline
Features & Explanation & Feature count\\
\hline\hline
Instruction related & Instruction count per category, register statistics (e.g., register reuse distance), operand histogram, etc. & 61\\
\hline
Memory access related & Code and data reuse distance & 48\\
\hline
Branch related & Count of branch taken/not-taken and branch entropy \cite{li24_sc} & 7\\
\hline
Remaining features & Interrupts, page faults, etc. & 12\\
\hline\hline
\end{tabular}
\label{features_breakdown}}
\end{table}

We experimented with two different dataset splits.
In one split, we used an 80/20 training/testing split, meaning 80\% of the dataset is used in training and the remaining 20\% is used for validation/testing. In other split, we deliberately split the dataset in such a way that it helps evaluating \scheme\ for unseen benchmarks.
We reorganized our dataset as follows: as unseen benchmarks, we hold out XZ, WRF, MCF, nab, cactubssn, and xalanbmk for 15 SKUs as our testing set. We selected these three benchmarks to represent core bound, uncore bound, and memory bound behavior. The rest of the benchmarks are used as our training set.


\subsection{Model Analysis}
\label{sec-model-analysis}
To develop a suitable model for \scheme, we experimented with simple as well as complex model architectures. We started with a simple {\em Linear Regression} model followed by a multi-layer perceptron model (MLP). Our initial results for both of these models showed unimpressive results, 
with our MLP only achieving a 0.45/0.78 train/test MSE. Therefore, we moved on to LSTM-based architectures.



\begin{figure}[ht]
\centering
\includegraphics[width=3.2in]{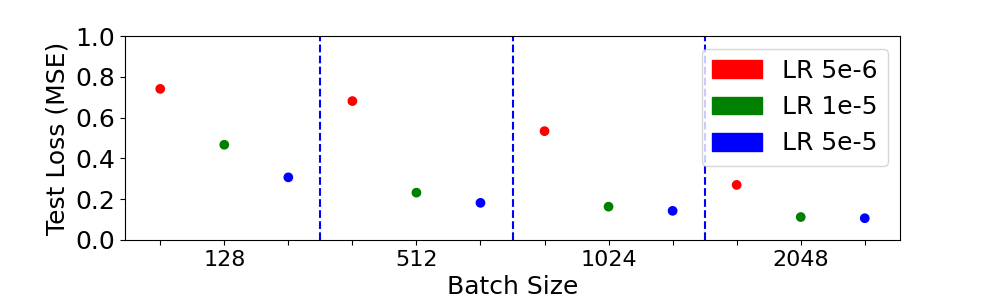}
\caption{Hyperparameter tuning of simple LSTM-based model.}
\label{fig-htune}
\vspace{-0.2cm}
\end{figure}

We experimented with a simple LSTM-based model first. This model consists of 
a two-layer LSTM with two fully connected layers. We searched a subset of the hyperparameter space by changing the learning rate and batch size. The results are shown in Figure~\ref{fig-htune}. Each model is trained for 10k epochs.
We see that there's a large spread between the best and worst performing parameter models. Notably, we see that the best performing models all have a batch size of 2048 and the worst performing models all have a batch size of 128.

\vspace{-0.2in}

\begin{figure}[ht]
\centering
\includegraphics[width = 0.7\columnwidth]{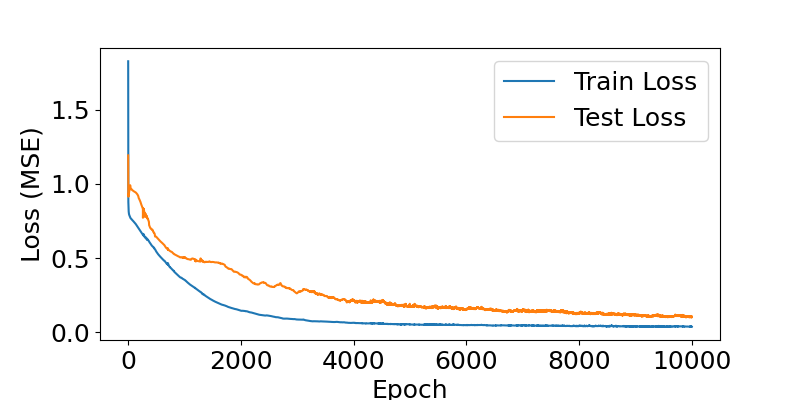}
\caption{Training and testing curve for the best performing model.}
\label{lstm_training}
\vspace{-0.2cm}
\end{figure}

Our best model achieves a training loss of 0.038 and a testing MSE of 0.100.
We further analyzed the train/test curve of the best model in order to evaluate whether or not is overfitting. In the train/test curve (Figure~\ref{lstm_training}), we see that the training and testing loss are both consistently improving throughout the entire training process. The lack of any divergence between the training and testing loss indicates that no overfitting has occurred and the model can stand to improve with further training epochs.

\vspace{-0.2in}


As we continued with our custom data split to evaluate unseen benchmarks, the simple LSTM-based model failed to converge. So, we developed and experimented with a hierarchical LSTM-based model (Figure~\ref{hier-model}). After conducting the same hyperparameter tuning process as with the single LSTM, we trained our best model on the custom dataset: achieving a train/test MSE of 0.010/0.029.

\begin{figure}[ht]
\centering
\includegraphics[width=0.7\columnwidth]{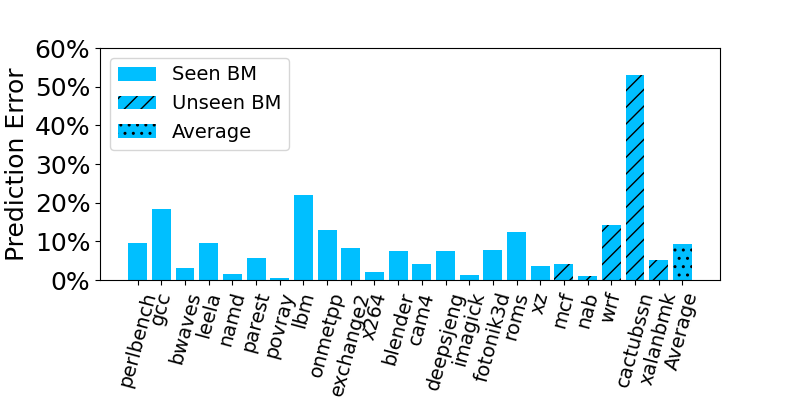}
\caption{\rev{Full benchmark IPC predictions using hierarchical LSTM model.}}
\label{wl_preds}
\vspace{-0.2cm}
\end{figure}

Figure~\ref{wl_preds} illustrates the average IPC prediction error of \scheme\ using the hierarchical LSTM model 
compared to the IPC obtained from silicon.
\scheme's predictions have an average absolute error of only 
\rev{9.35\%} which demonstrates that our model is capable of predicting IPC values fairly accurately for entire benchmarks despite being trained on fine-grained intervals.
Notably, when just considering seen/unseen benchmark predictions, we see that our seen benchmarks have an average error of 
\rev{7.64\%} and our unseen benchmarks have an average error of 
\rev{15.5\%}. \rev{
In our results, we see a few outliers like gcc, lbm and cactubssn, which have a prediction error significantly over 10\%. For cactubssn we observed that with higher core count (64x) IPC went down (0.18x) significantly in Silicon due to high L2 (5x) and LLC (5x) miss latency. It seems \scheme's accuracy degrades in predicting a lower silicon IPC value probably due to not learning the underlying reasons such as the high L2, LLC miss latency. We intend to further study what causes these outliers to have such poor performance and address such causes in our model architecture and training dataset in the future. Thus the results are similar or slightly degraded compared to PerfVec. Despite that, \scheme\ has an edge over PerfVec by not relying on simulation-based data.} The improvement in our hierarchical LSTM design over the single LSTM is partly due to the two independent LSTMs for handling uAIMs and hardware configurations. These feature sets display different properties, so allowing the model to train the weights for uAIMs and HW configs separately leads to better accuracy for our 15 SKUs.

\begin{table}[htpb]
\caption{Comparison of prediction time. \label{tab:table1}}
\centering
\scalebox{0.65}{
\begin{tabular}{||p{1.5cm}||p{1.5cm}||p{2.5cm}||p{2.5cm}||}
\hline\hline
\multirow{3}{*}{Scheme} & \multicolumn{3}{c||}{Prediction time (s)}\\
\cline{2-4}
 & & Full benchmark & Full Benchmark suite\\
 & 10B inst. & (3T inst. per BM) & (65T inst.) \\
\hline\hline
{\bf \scheme} & \textbf{1} & \textbf{9.91} & \textbf{228.91} \\
\hline
TAO~\cite{tao} & 5076 & 1.45$\times 10^6$ & 3.33$\times 10^8$ \\
\hline
SimNet \cite{simnet} & 6948 & 1.98$\times 10^6$ & 4.56$\times 10^8$\\
\hline\hline
\end{tabular}}
\label{timing}
\vspace{-0.4cm}
\end{table}

\subsection{Timing Analysis}
\label{sec-timing}

\scheme\ completes inference for the SPEC 2017 benchmark suite in about 2 minutes and 57 seconds, averaging 1 second per 10 billion instructions. This is {\em three orders} of magnitude faster than the state-of-the-art simulators like TAO~\cite{tao} and SimNet~\cite{simnet} which take 5076 seconds and 6948 seconds, respectively. This is because TAO and SimNet process instructions individually, while \scheme\ operates at the 10 million instruction granularity. \scheme's fast inference speed enables it to run the full benchmark of approximately 3 trillion instructions for multiple iterations, with each iteration taking only about 10 seconds to predict the IPC. For TAO and SimNet, we estimated the prediction time for the full benchmark and the entire benchmark based on their time to predict 10 billion instructions. Table~\ref{timing} summarizes this prediction time comparison. \rev{Notably, PerfVec outperforms our model in benchmark-wise prediction time; however, the paper doesn't include enough information to extrapolate this to the 10B instruction or full benchmark suite level}.

\section{Conclusion \& Future Work}
\label{sec-conc}
We started this project by asking if AI/ML can be used to augment the performance estimation step of SoC design cycle by supplementing detailed simulators with a fast and accurate alternative capable of predicting full benchmark performance.
Towards that end, we proposed \scheme, the {\em first} technique to accurately predict full benchmark performance without relying on detailed simulation or instruction-wise encoding. At the heart of \scheme\ is a hierarchical LSTM-based model that takes a trace of uAIMs from a program execution and predicts performance metrics. We presented the detailed design, implementation and evaluation of \scheme. Our initial experiments showed that \scheme\ can achieve an average IPC prediction error of 
\rev{9.35\%} for SPEC CPU 2017 benchmark suite while taking only 2 min 57 sec for the entire suite. This prediction error is comparable to prior state-of-the-art techniques~\cite{tao, li24_sc} while requiring 3 orders of magnitude less time. Our future work includes generalizing PAI across multiple benchmark suites and processor architectures spanning multiple generations.


\bibliographystyle{IEEEtran}
\bibliography{ref}

\vfill

\end{document}